\documentclass[useAMS,usenatbib,usegraphicx]{mn2e}
\begin{document}
\def \th {\thinspace}
\def \arcmin {\hbox{$^\prime$}}
\def \arcsec {\hbox{$^{\prime\prime}$}}
\def \chisq {$\chi ^{2}$}
\def\approxgt{\mathrel{\hbox{\rlap{\lower.55ex \hbox {$\sim$}} \kern-.3em \raise.4ex \hbox{$>$}}}}
\def\lesssim{\mathrel{\hbox{\rlap{\lower.55ex \hbox {$\sim$}} \kern-.3em \raise.4ex \hbox{$<$}}}}
\def\approxlt{\mathrel{\hbox{\rlap{\lower.55ex \hbox {$\sim$}} \kern-.3em \raise.4ex \hbox{$<$}}}}
\def \degmark {^\circ}
\def \sun {\hbox {$\odot$}}

\title[Hard X-ray features in the hotspots of Cyg\th A]
{Discovery of hard X-ray features around the hotspots of Cygnus\th A}

\author[Ba\l uci\'nska-Church et al.]
{M. Ba\l uci\'nska-Church$^{1 \, , \, 2}$, M. Ostrowski$^{2, 3}$, \L. Stawarz$^{2}$ and M. J. Church$^{1 \, , \, 2}$
\thanks{E-mail: mbc@star.sr.bham.ac.uk, mio@oa.uj.edu.pl,
stawarz@oa.uj.edu.pl, mjc@star.sr.bham.ac.uk} \\
$^1$School of Physics and Astronomy, University of Birmingham,
Birmingham B15 2TT, UK \\
$^2$Obserwatorium Astronomiczne, Uniwersytet Jagiello\'nski,
ul. Orla 171, 30-244 Krak\'{o}w, Poland \\
$^3$ Institut f\"ur Theoretische Physik IV, Ruhr-Universit\"at Bochum, D-44780 Bochum, Germany}

\date{Received June 22nd, 2004. Accepted November 8th, 2004.}

\pagerange{\pageref{firstpage}--\pageref{lastpage}} \pubyear{2004}

\maketitle

\label{firstpage}

\begin{abstract}
We present results of analysis of a {\it Chandra} observation of Cygnus\th A in which the X-ray hotspots 
at the ends of the jets are mapped in detail. A hardness map reveals previously unknown structure in the 
form of outer and inner hard arcs around the hotspots, with hardness significantly enhanced compared with 
the hotspot central regions. The outer hard arcs may constitute the first detection of the bow shock; the 
inner hard arcs may reveal where the jets impact on the hotspots. We argue that these features cannot result 
from electrons radiating by the synchrotron self-Compton process. Instead we consider two possible sources 
of the hard emission: the outer arcs may be due to thermal radiation of hot intracluster gas compressed at 
the bow shock. Alternatively, both outer and inner arcs may be due to synchrotron radiation of electrons accelerated in 
turbulent regions highly perturbed by shocks and shear flows. Comparison of measured hardness 
ratios with simulations of the hardness ratios resulting from these processes show that it is more difficult 
to explain the observations with a thermal model. Although we cannot rule out a thermal model, we argue
in favour of the non-thermal explanation. The hard regions in the secondary hotspots suggest that 
jet activity is still powering these hotspots.
\end{abstract}


\section{Introduction}

In the brightest radio galaxies, hotspots at the ends of the radio jets are often found which
radiate from radio to X-rays. Blandford \& Rees (1974) and Scheuer (1974) proposed
that these are formed where the jet interacts with the ambient medium and decelerates
abruptly forming a double shock structure. The forward shock (the `bow shock') compresses and
heats the intergalactic medium. The reverse shock (the `Mach shock') propagates within the jet,
converting the jet bulk kinetic energy into cosmic ray particles and magnetic field energy. The
interaction of the jet with the compressed ambient gas turns the shocked jet matter backwards to
expand into the radio lobes.  The observed hotspots are generally thought to be located downstream
of the reverse shock, but whether they extend to the bow shock is not clear.

Radio observations of the hotspots show highly polarized broadband emission of power law 
spectral form, without doubt synchrotron in origin. This emission extends in some cases to the 
optical band (Meisenheimer et al. 1989). The observations often reveal a complex
radio morphology in the jet terminal regions, with the presence of multiple hotspots (e.g. 
Hardcastle et al. 1998). Several explanations have been proposed for these, including formation
by precession of the jets on relatively short timescales (Scheuer 1982), or the deflection of the jets
at the terminal point due to extended oblique shocks at the jet head (Williams \& Gull 1985). The
time-dependent nature of the hotspots may also be related to modulation of the kinetic power
of the jet (Leahy \& Perley 1995; see also Stawarz et al. 2004).

As the synchrotron lifetime of the radio and optical emitting electrons (in the equipartition magnetic
field) is typically much shorter than the light-crossing time between the nucleus and the hotspot region
(Hargrave \& Ryle 1974, Meisenheimer et al. 1989, Brunetti et al. 2003), it is thought that there must
be {\it in situ} particle acceleration within the hotspots. Heavens \& Meisenheimer (1987) proposed a
particular model in which such an acceleration takes place at the Mach shock, providing a power law
population of ultrarelativistic electrons which cool radiatively by the synchrotron and inverse-Compton
processes in a localized downstream region where the magnetic field strength is high (the `continuous
injection' model). In most cases, the radio-to-infrared spectra of the hotspot emission can be explained
by this model (Meisenheimer et al. 1989).
Apart from this localized acceleration, some infrared and optical observations provide evidence for
an acceleration process distributed around the hotspots. Diffuse infrared emission
extending a few kpc 
beyond the hotspot was detected in 3C 445 (Prieto, Brunetti \& Mack 2002), Pictor A (R\"oser \& 
Meisenheimer 1987) and in 3C 273 (Meisenheimer 2003). It was suggested that the appropriate
mechanism consists of stochastic acceleration processes at strong turbulence surrounding the
hotspot (Meisenheimer, Yates \& R\"oser 1997, Prieto et al. 2002).

X-ray emission from several radio hotspots was detected using {\it ROSAT}. More recently, the
spatial resolution and high sensitivity of {\it Chandra} has allowed the study of a larger number of
hotspots in powerful radio sources (Hardcastle et al. 2004).
For some of these hotspots, the X-ray emission is consistent with production by the
synchrotron self-Compton process (SSC). This allows estimation of the magnetic field strength often
yielding values close to the equipartition value. However, X-ray hotspot emission in several sources
cannot be explained in terms of a SSC model (Hardcastle et al. 2004), suggesting that the keV
photons most probably have synchrotron origin.

In Cyg\th A (Fig.~1), the jets form primary hotspots B and E and secondary hotspots A and D in 
the Western and Eastern lobes, respectively, as designated by Hargrave \& Ryle (1974). The radio
hotspots A and D are highly polarized, roughly spherically shaped, with radius $\approx$ 2 kpc and
bolometric luminosity $L_{\rm syn} \approx 4 \times 10^{44}$ erg s$^{-1}$ (Carilli \& Barthel
1996). Carilli et al. (1991, 1999) and Meisenheimer et al. (1997) showed that the synchrotron
emission of hotspots A and D extending to at least $10^{12}$ Hz, with a high-frequency cut-off
at $\sim 10^{13}$ Hz implied by the optical upper limit, is consistent with the `continuous injection'
model. 
The breaks in the radio spectra allow physical parameters to
be constrained for these two hotspots, with the magnetic field strength derived
being consistent with the equipartition value $B_{\rm eq} \approx
3 \times 10^{-4}$ G, the related minimum hotspot pressure
$p_{\rm eq} \approx 3 \times 10^{-9}$ dyn cm$^{-2}$, and the
downstream velocity of the radiating plasma $\beta_{\rm out}
\approx 0.05 - 0.1$.

The hotspots A and D were detected in X-rays using {\it ROSAT} (Harris, Carilli \& Perley 1994),
who showed that the X-ray emission was non-thermal, as a thermal model required a particle number
density far above the upper limit allowed by the lack of internal radio depolarization, $n_{\rm th} 
< 4 \times 10^{-4}$ cm$^{-3}$ (Dreher, Carilli \& Perley 1987).  Spectral analysis of the hotspots
in the present {\it Chandra} observation allowed Wilson, Young \& Shopbell (2000) to obtain a X-ray
power law energy index $\alpha_{\rm X} \approx 0.8$, and a X-ray hotspot luminosity, $L_{\rm X}
\approx 10^{42}$ erg s$^{-1}$. These can be explained as SSC emission for a magnetic field strength
$B \sim 1.5 \times 10^{-4}$~G (Harris et al. 1994; Wilson, Young \& Shopbell 2000), a value close
to $B_{\rm eq}$. Because of this, the hotspots A and D in Cygnus\th A are generally accepted as
standard examples of astrophysical objects justifying the equipartition hypothesis (cf. Kino \& 
Takahara 2004). In the present paper, we find previously-unknown hardness variations across the X-ray 
hotspots of Cygnus\th A implying significant substructure.

\section{Observations and analysis}

We have analysed the observation of Cygnus\th A made with {\it Chandra} on May 21, 2000 UT
03:13:30 -- 13:25:55, with a total usable exposure of $\sim$34000 s. Data analysis was carried out using 
{\sc ciao} version 3.0.2 and the latest calibration data (Feb 2, 2004). Events files were produced which 
had been checked for bad pixels and high background, and corrections made for charge transfer inefficiency 
and gain variations in the image. X-ray images were produced using {\sc dmcopy} of size 350$\times$287
0.49$\arcsec$ pixel, sufficient to contain all of Cyg\th A. The intensity images were not converted
to flux images as this process makes assumptions about the spectrum, and the intensity image in the band 
0.3--8.0 keV image with 5~GHz radio contours superimposed is shown in Fig.~1.

\begin{figure}							
\caption{{\it Chandra} ACIS X-ray intensity image of Cygnus\th A in the energy band 0.3 -- 12 keV, with
5~GHz radio contours superimposed (kindly provided by C. L. Carilli).}
\label{fig1}
\end{figure}

We attempted to carry out spectral fitting by dividing each hotspot into an outer region (away from the 
nucleus) and an inner region. However, the total count in each of these regions was 140 which does not
allow sensible spectral fitting, not allowing discrimination between thermal and non-thermal models.
Previous analysis of this observation (Wilson et al. 2000) used the whole of hotspots A and D.

\begin{figure*}							
\caption{X-ray hardness with the intensity contours superimposed. Left panel: the counter jet hotspot D,
right panel: the jet hotspots A and B. Hardness increases from blue to yellow. Image sizes are 
$19^{\prime\prime} \times 15^{\prime\prime}$. The variation in hardness across the hotspots
is seen.}
\label{fig2}
\end{figure*}

Because of the difficulty of spectral fitting with smaller regions, we used X-ray hardness ratios to 
constrain spectral changes within the source. We obtained images using {\sc dmcopy} in two
energy bands: 2.0--8.0 keV and 0.3--2.0 keV, which were divided to give a hardness map. As there is
a wide range of intensity at the hotspots with pixels in outer regions having large Poisson
fluctuations, it is necessary to use either adaptive smoothing or adaptive binning (Sanders \& Fabian
2001). Firstly, we smoothed adaptively in each energy band using {\sc csmooth}. The hardness map
displays significant changes across the whole image of Cyg\th A: in the central plane perpendicular  
to the jets, along the jets and in the hotspots. In this Letter we discuss the hardness variations in the 
hotspots. Care has to be taken in using adaptive smoothing because the Gaussian width $\sigma$ used in 
smoothing the high energy band with less counts will be larger than in the low band, and this {\it might} 
lead to a hard annulus around a hotspot as an artifact. To avoid this possibility we carried out the
smoothing in both bands using $\sigma$ values at every position in the image derived from the band 
0.3--2.0 keV. The resulting hardness map for each hotspot is shown in Fig.~2, with X-ray intensity contours  
(0.3--8.0 keV) superimposed. Strong hardness variations are seen in each hotspot, and clearly the hardness 
ratio (HR) does not correlate with X-ray intensity. On the contrary, the peak of X-ray intensity is 
the least hard, with hard features around the outsides which, being extended, we term hard arcs.
Spot A in particular shows the softer
region extending along the axis of elongation of the intensity contours, approximately perpendicular to
the direction of the jet. The HR values increase by a factor of two to three from the centres to
the hard outer regions. A possible objection to adaptive smoothing is that it involves some
redistribution of the counts. We also carried out adaptive binning of the data in the two energy bands
to avoid this, which bins together several pixels in the outer hotspots to
reduce Poisson fluctuations to a specified level (35 percent). There was general agreement with
the results of adaptive smoothing, with the centres of the hotspots being less hard, and hard regions
around the hotspots. Beyond this, the count statistics do not allow more precise definition of
the hard region geometry. We noted a slight tendency for the hard regions to be moved radially outwards
in adaptive smoothing probably related to the redistribution of counts. However, the adaptively smoothed
images are better for display.

We used the adaptively binned images to obtain mean values of hardness since this is equivalent to
using raw data. Average values of hardness were obtained from 4 rebinned pixels (14 primitive pixels)
for spot A including the hard regions on both sides of the hotspot, giving HR = 0.77 $\pm$ 0.16. The central 
region had HR = 0.29 $\pm$ 0.13. Thus the change is highly significant. For spot D, the corresponding 
values are HR=0.79 $\pm$ 0.16 in 9 pixels and 0.33$\pm$0.10., and for spot B the values are 0.79 $\pm$ 0.16
and 0.32 $\pm$ 0.11.

To assess the implications of the hardness changes, spectral data were simulated 
for a simple absorbed power law: {\sc ab}$*${\sc pl} for a range of photon power law index $\Gamma$,
and for an absorbed Mekal model: {\sc ab}$*${\sc me} for a range of $kT$, 
using large normalizations to avoid Poisson fluctuations. Then the data were divided into the same two 
energy bands as in our analysis and the hardness ratio found. The column density was fixed at $0.3\times 
10^{22}$ atom cm$^{-2}$ from radio determinations near the hotspot (Dickey 
\& Lockman 1990). The mean HR in the hard arcs (above), and at peak intensity in each hotspot
are compared with the simulations (dashed curves) in Fig.~3.
The X-ray intensity falls sharply with radial distance from the centre of each hotspot, continuing to decrease 
on the outside of each hotspot (away from the nucleus). We chose to determine the background  
at an outside radial position of 3.5$\arcsec$, but this may still be an overestimate. Because of 
this uncertainty, in Fig.~3 we plot hardness values {\it before} background correction and show the probable
maximum extent of the correction for the hard arc points as dotted lines. The correction would {\it increase}
the change in hardness across each hotspot. At the peak of X-ray
intensity, the effect of background subtraction is negligible. 

The above HR values for the hard arcs intersect the power law curve at 
$\Gamma$ = 1.23 $\pm$ 0.20, 1.20 $\pm$ 0.20 and 1.20 $\pm$ 0.20 for spots A, D and B,
and for the peak intensity at $\Gamma$ = 2.18 $\pm$ 0.42, 2.07 $\pm$ 0.30 and 2.10 $\pm$ 0.34
for the same three hotspots.
Thus the change in power law index between the peak intensity and the hard arc is $\Delta\Gamma$ = 
0.96 $\pm$ 0.47,  0.88 $\pm$ 0.36 and 0.91 $\pm $ 0.39 for hotspots A, D and B. 

When compared with the Mekal model, it can be seen that hardness 
ratios of 0.29, 0.33 and 0.32 can be {\it formally} explained by a $kT$ of 2.85, 3.22 and 3.09 keV 
at the peak of X-ray intensity, although we know in this case that the emission is non-thermal. The 
hard arcs require $kT$ = 96 (-84), 180 (-165) and 177 (-163) keV for spots A, D and B;
at the upper limit of HR, there is no intersection with the curve.

\vskip 2mm
\noindent {\bf
3\hskip 4mm ORIGIN OF THE HARD FEATURES}

\vskip 1 mm\noindent
The hard arcs discovered in the present work cannot be explained in the framework of the SSC process. 
Carilli et al. (1991) argued that at the edges of the hotspots the shock-generated electron energy distribution 
would adopt the Jaffe-Perola or the Kardashev-Pacholczyk shape, i.e. a sharp cut-off or steep power law in the
high energy range. Thus SSC X-ray emission would always be (at least a bit) softer at
the outer edge of the hotspot than inside at the X-ray intensity peak,
contrary to Fig.~2. Thus, we consider other possible mechanisms for producing the hard regions.

\begin{figure*}
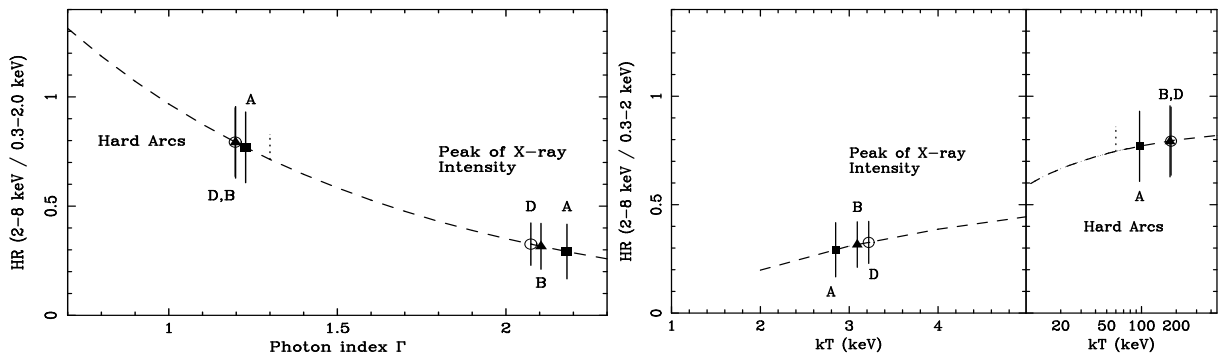
                                         
\includegraphics[width=46mm,height=80mm,angle=270]{fig3a.ps}        
\includegraphics[width=46mm,height=80mm,angle=270]{fig3b.ps}        
\caption{Comparison of measured enhancements of hardness in the hard arcs with
simulations. Left panel: non-thermal model; right panel: thermal model. The dotted lines
show the maximum correction due to background subtraction. In the thermal model, the point 
for hotspot A never intersects the dashed line, and is arbitrarily shown at 50 keV; 
the point for hotspot D intersects at 38 keV.}
\label{fig3}
\end{figure*}

A natural source of hardening for the outer arcs, i.e. on the outsides of the hotspots,
could be the thermal radiation of ambient intracluster gas compressed and 
heated by the bow shock. This shocked gas forms the sheath behind the bow shock in pressure equilibrium
with the shocked jet plasma, for which the minimum pressure is $p_{\rm eq} \approx 3 \times 10^{-9}$
dyn cm$^{-2}$ (Sect.~1). Ram pressure confinement of the hotspot implies that $\rho_{\rm g} \, v_{\rm sh}^2
= p_{\rm eq}$, and hence that $\mathcal{M}_{\rm sh} = (p_{\rm eq} / \hat{\gamma}_{\rm g} \, p_{\rm g}
)^{1/2}$, where $\rho_{\rm g}$, $p_{\rm g}$ and $\hat{\gamma}_{\rm g}$ are the density, the pressure and
the ratio of specific heats of the unperturbed ambient gas (`g'). $v_{\rm sh}$ and $\mathcal{M}_{\rm sh}$ are
the bow shock velocity and Mach number: $\mathcal{M}_{\rm sh} \equiv v_{\rm sh} /
c_{\rm s, \, g}$, where the sound speed of the unperturbed ambient gas is $c_{\rm s, \, g} = 
(\hat{\gamma}_{\rm g} \, p_{\rm g} / \rho_{\rm g})^{1/2}$. For $\hat{\gamma}_{\rm g} = 5 / 3$ and 
$p_{\rm g} \approx 10^{-10}$ dyne cm$^{-2}$
(Smith et al. 2002) one gets $\mathcal{M}_{\rm sh} \approx 5$ which is
a lower limit as the \emph{minimum} hotspot pressure was used. From the Mach number, we can estimate the
temperature of the shock-heated gas ($kT_{\rm g, \, +}$) and the compression ratio of the bow shock from the
Rankine-Hugoniot conditions, giving $kT_{\rm g, \, +} \approx 8.7 \times kT_{\rm g} \approx 43.4$ keV and
$n_{\rm g, \, +} \approx 3.6 \times n_{\rm g} \approx 0.036$ cm$^{-3}$ for a temperature and density of the
unperturbed intracluster gas of $kT_{\rm g}$ $\approx 5$ keV and $n_{\rm g} \approx 10^{-2}$ cm$^{-3}$
(Smith et al. 2002; see also Harris et al. 1994). The hardness of the hard outer arcs in spots A and  
D (Fig.~2) correspond to $kT$ $\sim$96 and 180 keV in the simulated data with large errors, i.e. $kT$ $>$
10 and 15 keV, respectively. Thus we 
cannot rule out thermal origin of the hard outer arcs. However, the plasma density
at the inner hard regions probably close to the jet terminal shock is thought to be low so that thermal radiation
cannot explain these.

A thermal origin of the outer arcs would be supported by observation of related structures 
in the Faraday rotation radio maps, because the compression behind the bow
shock can cause some change in polarization of a fraction of the radio emission 
(Carilli, Perley \& Dreher (988). We evaluate the total Faraday rotation  
within the shocked thermal gas layer using the above shock and intracluster 
parameters, i.e. an electron density in the shocked sheath $n_{\rm g, \, +} \approx 0.04$ cm$^{-3}$
and a mean magnetic field $B_{\rm g, \, +} \approx 20$ $\mu$G. The  sheath thickness $h$ is the
hotspot size divided by $2\sqrt{5}$ (Carilli et al. 1988), i.e. $\approx 200$ pc. As the 
sheath is curved, resembling the Earth's bow shock in the solar wind, the mean radio wave propagation 
length, $l$, through this layer will be a few times $h$, say 500 pc. For a wavelength $\lambda$
= 6 cm and $B_{\rm g, \, \|} \approx B_{\rm g, \, +}/2$, with these units, the characteristic Faraday 
rotation angle $\Delta \Psi = 8.1 \times 10^{-3} \lambda^2 B_{\rm g, \,\|}\, l\, n_{\rm g, \, +}$ is
$\Delta \Psi \approx 0.6$ rad. Thus {\it variation} in the rotation angle across the region
at $\lambda \lesssim 6$ cm will also be small, and so depolarization will be small.

Both inner and outer hard regions may be non-thermal in origin, but the extremely flat spectral index 
$\alpha_{\rm X} \sim 0$ implied 
by $\Gamma$ $\sim$ 1 excludes inverse Compton radiation produced by the electrons responsible for the 
hotspot radio emission, as the radio spectral index in the hotspot regions is $\alpha_{\rm R}$ $\approxgt$ 
0.5 Instead, we have to consider synchrotron emission of an additional, high-energy flat electron population 
$n_{\rm e}(\gamma) \sim \gamma^{-1}$, where $\gamma$ is the electron Lorentz factor. A natural 
possibility for the formation of such an additional electron component is turbulent particle acceleration within 
an extended region where the particle escape/advection timescale is much longer than the radiative 
timescale (Stawarz \& Ostrowski 2002; Stawarz et al. 2004 and references therein). 
For the outer arcs, this could be near the contact discontinuity; for the inner arcs near the 
jet terminal shock. As suggested by 
numerical simulations, the hotspot downstream regions near the contact discontinuity between shocked jet 
fluid and shocked ambient gas can be indeed highly turbulent (Mizuta, Yamada \& Takabe 2004
and de\th Young 2002). The electrons in such regions are subject to 
repeated scattering by fast moving MHD waves and undergo diffusion in momentum space. 
This process may produce cosmic ray electrons with spectrum 
determined by the acceleration timescale, $t_{\rm acc}(\gamma)$, the radiative 
loss timescale, $t_{\rm rad}(\gamma)$, and the timescale for particle escape from the acceleration region, 
$t_{\rm adv}(\gamma)$. For $t_{\rm rad} \ll t_{\rm adv}$ the accelerated electrons pile up below the 
maximum energy $\gamma_{\rm max}$ given by the equality $t_{\rm rad} = t_{\rm acc}$, forming a very 
flat energy distribution for $\gamma \approxlt \gamma_{\rm max}$, as illustrated for solar flare 
conditions by Petrosian \& Donaghy (1999). The efficiency and the characteristics of the process in the
hotspots are unknown, but the energy density of the electrons, $u_{\rm e}$, cannot
exceed the energy density of the turbulent magnetic field, $u_{\rm T}$.

For hotspots A and D, using parameters derived from radio spectra (Sect. 1)
the timescale for particles to escape from the acceleration site considered is $t_{\rm 
adv} \sim L / \beta_{\rm out} c = 3.3 \times 10^4$ yr, where the hotspot size $L \sim 1$ kpc, and we take
$\beta_{\rm out} \approx 0.1$. The 
radiative (synchrotron) timescale is $t_{\rm rad} \sim 6 \pi \, m_{\rm e} c / \sigma_{\rm T} \, B^2 \, \gamma 
= 2.5 \times 10^9 \, \gamma^{-1}$ yr with $B \sim 10^{-4}$ G. Thus, for high energy electrons with 
$\gamma$ $\approxgt $ 10$^5$, $t_{\rm rad} < t_{\rm adv}$, as required for formation of the flat-spectrum 
high energy electron distribution. For the particle acceleration timescale, we take the `optimistic' estimate 
(Stawarz \& Ostrowski 2002): $t_{\rm acc} \sim \zeta^{-1} \, (c / v_{\rm sc})^2 \, (r_{\rm e} / c) = 2.1 \times 
10^{-11} \, \zeta^{-1} \, \beta_{\rm sc}^{-2} \, \gamma$ yr, where $r_{\rm e}$ = $\gamma \, m_{\rm e} c^2 
/ e \, B$ is the electron gyroradius, $v_{\rm sc}$ = $\beta_{\rm sc} c$ is a characteristic velocity of the 
scattering centres (propagating magnetic field inhomogeneities, weak oblique shocks, etc.) and $\zeta$ = 
$u_{\rm T} / u_{\rm B}$ is the ratio of the energy density of the turbulent magnetic field to that of 
the large-scale mean field.
$\zeta$ is expected to be less than unity. $v_{\rm sc}$ can be identified with 
the Alfv\'en velocity or with the sound speed in the downstream region. The jet plasma is mildly relativistic
and so the sound speed in the acceleration region can reach high values $\sim c/\sqrt{3}$, and for approximate
equipartition between the downstream particles and the downstream magnetic field, the Alfv\'en speed is
expected to be in the range $\sim 0.1 - 1 \, c$. Therefore we conservatively take $\beta_{\rm sc} \sim 0.1$. The maximum
electron Lorentz factor and the maximum photon energy of the resulting synchrotron emission are then
$\gamma_{\rm max} \sim 10^{10} \, \zeta^{1/2} \, \beta_{\rm sc}$ and $\varepsilon_{\rm max} \sim 100 \,
\zeta \, \beta_{\rm sc}^2$ MeV, so that for $\zeta = 0.1$ and $\beta_{\rm sc} = 0.1$, $\gamma_{\rm max}
\sim 3 \times 10^8$ and $\varepsilon_{\rm max} \sim 100$ keV.

The synchrotron luminosity of the flat electron population can be estimated as $L_{\rm syn} \sim 10^{38} \,
V \, \eta \, \gamma_{\rm max}$ erg s$^{-1}$ (e.g. Stawarz et al. 2004), where $V$ is the volume
in kpc$^3$ and $\eta$ = $u_{\rm e} / u_{\rm B}$ is expected to be much less
than unity for this electron component. Thus, we conclude that the synchrotron luminosity of this population
of high-energy electrons \emph{can be comparable} with the SSC luminosity of the `primary' electrons
responsible for the hotspot radio emission.

\vskip - 2mm
\section{Discussion and Conclusions}

Our analysis has revealed outer and inner regions of enhanced hardness for the secondary hotspots. The outer hard arcs
may constitute a detection of the bow shocks. The inner hard arcs may indicate
the sites of the jet terminal shocks and suggest that the jets still interact with the secondary hotspots.
We note a similar inner feature in the primary hotspot B, 
suggesting that the jet is interacting with this hotspot to form a jet terminal shock in the
synchrotron lobe formed by earlier jet activity. The absence of an outer arc
may be explained by the jet not forming a bow shock until it reaches the intracluster plasma outside the lobe
(cf. Stawarz 2004).

As the present data do not allow detailed spectral analysis of the hard features in regions of 
relatively low X-ray brightness we do not attempt an elaborate physical model, but consider qualitatively 
two possible mechanisms. The first is thermal radiation of shocked plasma
at the bow shock. In this case, the present results would demonstrate the presence of hot 
gas in the plasma sheath between the bow shock and the contact discontinuity separating it from the compressed 
jet non-thermal medium. 
The free-free radiation of a shocked one-temperature plasma may explain the hardness of the hard outer
arcs in hotspot A and D and the temperature estimated for a shock heated region is
consistent with values implied by observation although these are not well-constrained.
The presence of such plasma could be revealed in high resolution Faraday rotation and/or depolarization maps
if observed at frequencies $\nu \lesssim 5 {\rm GHz}$. However, it is a problem for the thermal model
to explain the inner hard arcs because of their apparent location near the jet terminal shocks, where
thermal plasma is expected to be absent.

The second process is efficient second-order Fermi acceleration acting non-uniformly within the hotspots. 
The results would imply a highly perturbed medium in the sheared regions near the contact 
discontinuity and, possibly, near the jet terminal shock. Thus because it is reasonable to assume 
the same process forms both inner and outer arcs, and the thermal model cannot explain the inner arcs, 
we suspect that the hard features discovered in the present work are a manifestation of continuous particle 
acceleration taking place for the outer arcs close to the contact discontinuity between the downstream jet 
plasma and the shocked outer gas, and for the inner arcs near the jet terminal shock.

\thanks{We thank Dr. H. Ebeling for helpful discussions; MO thanks Prof. R. Schlickeiser for his hospitality during 
a stay in Bochum.This work made use of the {\it Chandra} data archive and was supported in part
by the Polish Committee for Scientific Research by grants PBZ-KBN-054/P03/2001 and KBN-1528/P03/2003/25.}

\end{document}